\ifx\pdfvariable\undefined
\pdfoutput=1
\fi
\AtBeginDocument{\paperdetails{
  submitted=2017-07-31,
  published=2017-12-06,
  year=2018,
  volume=2,
  issue=2,
  articlenumber=4,
}}
\documentclass[english]{programming}
\usepackage[backend=biber]{biblatex}
\usepackage{balance}
\usepackage{pdfpages}
\usepackage{pgfplots}
\usepgfplotslibrary{statistics}
\usepackage{amsmath}
\usepackage{pdfpages}
\usepackage{listings}
\usepackage{colortbl}
\usepackage{xcolor}
\usepackage{pdfpages}
\usepackage{textcomp}
\usepackage[hyphens]{url}
\usepackage{enumitem}

\makeatletter
\lstdefinelanguage{llvm}{
  morecomment = [l]{;},
  morestring=[b]",
  sensitive = true,
  classoffset=0,
  morekeywords={
    define, declare, global, constant,
    internal, external, private,
    linkonce, linkonce_odr, weak, weak_odr, appending,
    common, extern_weak,
    thread_local, dllimport, dllexport,
    hidden, protected, default,
    except, deplibs,
    volatile, fastcc, coldcc, cc, ccc,
    x86_stdcallcc, x86_fastcallcc,
    ptx_kernel, ptx_device,
    signext, zeroext, inreg, sret, nounwind, noreturn,
    nocapture, byval, nest, readnone, readonly, noalias, uwtable,
    inlinehint, noinline, alwaysinline, optsize, ssp, sspreq,
    noredzone, noimplicitfloat, naked, alignstack,
    module, asm, align, tail, to,
    addrspace, section, alias, sideeffect, c, gc,
    target, datalayout, triple,
    blockaddress
  },
  classoffset=1, keywordstyle=\color{purple},
  morekeywords={
    fadd, sub, fsub, mul, fmul,
    sdiv, udiv, fdiv, srem, urem, frem,
    and, or, xor,
    icmp, fcmp,
    eq, ne, ugt, uge, ult, ule, sgt, sge, slt, sle,
    oeq, ogt, oge, olt, ole, one, ord, ueq, ugt, uge,
    ult, ule, une, uno,
    nuw, nsw, exact, inbounds,
    phi, call, select, shl, lshr, ashr, va_arg,
    trunc, zext, sext,
    fptrunc, fpext, fptoui, fptosi, uitofp, sitofp,
    ptrtoint, inttoptr, bitcast,
    ret, br, indirectbr, switch, invoke, unwind, unreachable,
    malloc, alloca, free, load, store, getelementptr,
    extractelement, insertelement, shufflevector,
    extractvalue, insertvalue,
  },
  alsoletter={\%},
  keywordsprefix={\%},
}
\makeatother

\newcommand\code[1]{\texttt{#1}}

\begin{document}
\paperdetails{
perspective=art,
area={Security programming; Metaprogramming and reflection; Interpreters, virtual machines, and compilers}
}
\title{Introspection for C and its Applications to Library Robustness}
\author[a]{Manuel Rigger}
\authorinfo[photos/rigger]{
is a PhD student at Johannes Kepler University Linz, Austria.
His main research interests are low-level software security, run-time detection of bugs in C, and language implementation.
Currently, he works on the safe execution of low-level languages on the Java Virtual Machine.
Contact him at
\email{manuel.rigger@jku.at}.}
\author[a]{René Mayrhofer}
\authorinfo[photos/mayrhofer]{
heads the Institute of Networks and Security at Johannes Kepler University Linz, Austria, and the Josef Ressel Center on User-friendly Secure Mobile Environments (u'smile).
His research interests include computer security, mobile devices, network communication, and machine learning, which he brings together in his research on securing spontaneous, mobile interaction.
Contact him at
\email{rene.mayrhofer@jku.at}}
\author[b]{Roland Schatz}
\authorinfo[photos/schatz]{
is a senior researcher at Oracle Labs. His research interests are programming languages, virtual machines and dynamic compilation.
His current research focus is cross-language interoperability between dynamic and static languages.
He received his PhD in Computer Science from Johannes Kepler University in 2013.
Contact him at
\email{roland.schatz@oracle.com}.}
\author[b]{Matthias Grimmer}
\authorinfo[photos/grimmer]{
is a senior researcher at Oracle Labs.
His research focuses on compilers and virtual machines.
Grimmer received his PhD in Computer Science from Johannes Kepler University in 2015.
His PhD thesis is titled ``Cross-language interoperability in a multi-language runtime''.
Contact him at \email{contact@matthiasgrimmer.com}.}
\author[a]{Hanspeter M\"{o}ssenb\"{o}ck}
\authorinfo[photos/moessenboeck]{
is a professor of Computer Science at the Johannes Kepler
University in Linz, Austria, and the head of the Institute for System Software. He mainly
works on programming languages, compilers, and virtual machines with a special focus on
dynamic compiler optimizations. Other areas of interest include application performance
monitoring, software visualization, as well as static and dynamic program analysis.
Contact him at
\email{hanspeter.moessenboeck@jku.at}.}
\affiliation[a]{Johannes Kepler University Linz, Austria}
\affiliation[b]{Oracle Labs, Austria}


\authorrunning{M. Rigger, R. Mayrhofer, R. Schatz, M. Grimmer, and H. Mössenböck}

\maketitle
\begin{abstract}
\textbf{Context:}
In C, low-level errors, such as buffer overflow and use-after-free, are a major problem, as they cause security vulnerabilities and hard-to-find bugs.
C lacks automatic checks, and programmers cannot apply defensive programming techniques because objects (e.g., arrays or structs) lack run-time information about bounds, lifetime, and types.

\textbf{Inquiry:}
Current approaches to tackling low-level errors include dynamic tools, such as bounds or type checkers, that check for certain actions during program execution.
If they detect an error, they typically abort execution.
Although they track run-time information as part of their runtimes, they do not expose this information to programmers.

\textbf{Approach:}
We devised an introspection interface that allows C programmers to access run-time information and to query object bounds, object lifetimes, object types, and information about variadic arguments.
This enables library writers to check for invalid input or program states and thus, for example, to implement custom error handling that maintains system availability and does not terminate on benign errors.
As we assume that introspection is used together with a dynamic tool that implements automatic checks, errors that are not handled in the application logic continue to cause the dynamic tool to abort execution.

\textbf{Knowledge:}
Using the introspection interface, we implemented a more robust, source-compatible version of the C standard library that validates parameters to its functions.
The library functions react to otherwise undefined behavior; for example, they can detect lurking flaws, handle unterminated strings, check format string arguments, and set \emph{errno} when they detect benign usage errors.

\textbf{Grounding:}
Existing dynamic tools maintain run-time information that can be used to implement the introspection interface, and we demonstrate its implementation in Safe Sulong, an interpreter and dynamic bug-finding tool for C that runs on a Java Virtual Machine and can thus easily expose relevant run-time information.

\textbf{Importance:}
Using introspection in user code is a novel approach to tackling the long-standing problem of low-level errors in C.
As new approaches are lowering the performance overhead of run-time information maintenance, the usage of dynamic runtimes for C could become more common, which could ultimately facilitate a more widespread implementation of such an introspection interface.
\end{abstract}

\section{Introduction}
Since the birth of C almost 50 years ago, programmers have written many applications in it.
Even the advent of higher-level programming languages has not stopped C's popularity, and it remains widely used as the second-most popular programming language~\cite{tiobe}.
However, C provides few safety guarantees and suffers from unique security issues that have disappeared in modern programming languages.
Buffer overflow errors, where a pointer that exceeds the bounds of an object is dereferenced, are the most serious issue in C~\cite{bufferoverflows}.
Other security issues include use-after-free errors, invalid free errors, reading of uninitialized memory, and memory leaks.
Numerous approaches exist that prevent such errors in C programs by detecting these illegal patterns statically or during run time, or by making it more difficult to exploit them~\cite{pastpresentfuture,sok,runtime}.
When an error happens, run-time approaches abort the program, which is more desirable than risking incorrect execution, potentially leaking user data, executing injected code, or corrupting program state.

However, we believe that in many cases programmers could better respond to illegal actions in the application logic if they could use the metadata of run-time approaches (e.g., bounds information) to check invalid actions at run time and prevent them from happening.
Library implementers in particular could use it to protect themselves from user input and to compensate for the lack of exception handling in C.
For example, if they could check that an access would go out-of-bounds in a server library, they could log the error and ignore the invalid access to maintain availability of the system (as in failure-oblivious computing~\cite{failureobliviouscomputing}).
If the error happened in the C standard library instead, they could set the global integer variable \code{errno} to an error code, for example, to \code{EINVAL} for invalid arguments.
Furthermore, a \emph{special value} (such as \code{-1} or \code{NULL}) could be returned to indicate that something went wrong.
Finally, explicit checks could prevent lurking flaws that would otherwise stay undetected.
For example, in the case that a function does not actually access an invalid position in the buffer, bounds checkers cannot detect when an incorrect array size is passed to the function.
Using bounds metadata, programmers could validate the passed against the actual array size.

In this paper, we present a novel approach that allows C programmers to query properties of an \emph{object} (primitive value, struct, array, union, or pointer) so that they can perform explicit sanity checks and react accordingly to invalid arguments or states.
These properties comprise the bounds of an object, the memory location, the number of arguments of a function with varargs, and whether an object can be used in a certain way (e.g., called as a function that expects and returns an \code{int}).
The presented approach is \emph{complementary} to dynamic tools, and does not aim to replace them.
Programmers can insert custom input validations and error-handling logic where needed, but the dynamic tool that tracks the exposed metadata still aborts execution for errors that are not handled at the application level.
Ultimately, this provides programmers with greater flexibility and increases the robustness of libraries and applications, defined as ``[t]he degree to which a system or component can function correctly in the presence of invalid inputs or stressful environmental conditions''~\cite{glossary}.

As a proof of concept, we implemented the introspection interface for Safe Sulong~\cite{ecoopds}, a bug-finding tool and interpreter with a dynamic compiler for C.
Safe Sulong prevents buffer overflows, use-after-free, variadic argument errors, and type errors by checking accesses and aborting execution upon an invalid action.
It already maintains relevant run-time information that it can expose to the programmer.

In a case study, we demonstrate how the introspection functions facilitate re-implementing the C standard library (\code{libc}) to validate input arguments.
We use this \code{libc} in Safe Sulong as a source-compatible, more robust drop-in replacement for the GNU C Library.
In contrast to the GNU C Library and other implementations, it can detect lurking flaws, handle unterminated strings, check format string arguments, and -- instead of terminating execution -- set \emph{errno} when errors occur.

A plethora of other dynamic-bug finding tools and runtimes for C exist, and they could expose their run-time data via the introspection functions introduced in this paper.
For example, bounds checkers~\cite{softbound,asan,mudflap,libsafe} could expose bounds information.
Temporal memory safety tools~\cite{cets,msan,memcheck,purify,valgrind,memcheck,drmemory} could expose information about memory locations.
Variadic argument checkers~\cite{varargs} and type checkers~\cite{typesan,libcrunch} could expose information about variadic arguments and types.
There are also combined tools that, for example, provide protection against both out-of-bounds accesses and use-after-free errors~\cite{cets,softbound,managedc}.

As the overhead of dynamic tools is decreasing~\cite{softbound,cets,libcrunch,asan,msan}, they could become standard in production, similar to stack canaries and address space layout randomization~\cite{sok}.
At this point in time, a wider adoption of the introspection functions (or a subset thereof) seems feasible.
Additionally, we envisage that dynamic tools available now could distribute specialized libraries that benefit from introspection, as we will demonstrate using Safe Sulong's \code{libc} as an example.

In summary, this paper contributes in the following ways:
\begin{itemize}[leftmargin=10pt]
    \item{We present introspection functions designed to allow programmers to prevent illegal actions that are specific to C (Section~\ref{sec:api}).}
    \item{We demonstrate how we implemented the introspection functions in Safe Sulong, a bug-finding tool and interpreter with a dynamic compiler for C (Section~\ref{sec:implementation}).}
    \item{In a case study, we show how using introspection increases the robustness of the C standard library (Section~\ref{sec:casestudy}).}
\end{itemize}

\section{Background}
\label{sec:background}
In C, the lack of type and memory safety causes many problems, such as hard-to-find bugs and security issues.
Moreover, manual memory management puts the burden of deallocating objects on the programmer.
Consequently, C programs are plagued by vulnerabilities that are unique to the language.
Faults can invoke undefined behavior, so compiled code can crash, compute unexpected results, and corrupt or read neighboring objects~\cite{undefined, undefined2}.
It is often impossible to design C functions such that they are secure against usage errors, since they cannot validate passed arguments or global data.
Below we provide a list of errors and vulnerabilities in C programs that we target in this work.
\begin{description}[leftmargin=0pt]
    \item[Out-of-bounds errors.]{
        Out-of-bounds accesses in C are among the most dangerous software errors~\cite{top25,bufferoverflows}, since -- unlike higher-level languages -- C does not specify automatic bounds checks.
        Further, objects have no run-time information attached to them, so functions that operate on arrays require array-size arguments.
        Alternatively, they need conventions such as terminating an array by a special value.

        Listing~\ref{out-of-bounds} shows a typical buffer overflow.
        The \code{read\_number()} function reads digits entered by the user into the passed buffer \code{arr} and validates that it does not write beyond its bounds.
        However, its callee passes \code{-1} as the \code{length} parameter, which is (through the \code{size\_t} type) treated as the unsigned number \code{SIZE\_MAX}.
        Thus, the bounds check is rendered useless, and if the user enters more than nine digits, the \code{read\_number()} function overflows the passed buffer.
\begin{lstlisting}[language=C, float=b,label=out-of-bounds, caption=Passing \code{-1} to the \code{size\_t} parameter renders the range check useless and could cause an out-of-bounds error while writing read characters to \code{arr}]
void read_number(char* arr, size_t length) {
  int i = 0;
  if (length == 0) return;
  int c = getchar();
  while (isdigit(c) && (i + 1) < length) {
    arr[i++] = c; c = getchar();
  }
  arr[i] = '\0';
}
// ...
char buf[10];
read_number(buf, -1);
printf("%s\n", buf);
\end{lstlisting}

        A recent similar real-world vulnerability is CVE-2016-3186, where a function in libtiff cast a negative value to \code{size\_t}.
        As another example, in CVE-2016-6823 a function in ImageMagick caused an arithmetic overflow that resulted in an incorrect image size.
        Both faults resulted in buffer overflows.
    }
    \item[Memory management errors.]{Objects that are allocated in different ways (e.g., on the stack or by \code{malloc()}) have different lifetimes, which influences how they can be used.
        For example, it is forbidden to access memory after it has been freed (otherwise known as an access to a \emph{dangling pointer}).
        Other such errors include freeing memory twice, freeing stack memory or static memory, and calling \code{free()} on a pointer that points somewhere into the middle of an object~\cite{cets}.
        Listing~\ref{use-after-free} shows examples of a use-after-free and a double-free error.
        Firstly, when \code{err} is non-zero, the allocated pointer \code{ptr} is freed and later accessed again as a dangling pointer in \code{logError()}.
        Secondly, the code fragment attempts to free the pointer again after logging the error, which results in a double-free vulnerability.
    \begin{lstlisting}[language=C, float=tp,label=use-after-free,caption=Use-after-free error which is based on an example from the CWE wiki]
char* ptr = (char*) malloc(SIZE * sizeof(char));
if (err) {
  abrt = 1; free(ptr);
}
// ...
if (abrt) {
  logError("operation aborted", ptr); free(ptr);
}
// ...
void logError(const char* message, void* ptr) {
    logf("error while processing %p", ptr);
}
    \end{lstlisting}
        C does not provide mechanisms to retrieve the lifetime of an object, which would allow checking and preventing such conditions.
        Consequently, use-after-free errors frequently occur in real-world code.
        For example, in CVE-2016-4473 the PHP Zend Engine attempted to free an object that was not allocated by one of \code{libc}'s allocation functions.
        Other recent examples include a dangling pointer access and a double free error in OpenSSL (CVE-2016-6309 and CVE-2016-0705).
    }
    \item[Variadic function errors.]{
        Variadic functions in C rely on the programmer to pass a count of variadic arguments or a format string.
        Furthermore, a programmer must pass the matching number of objects of the expected type.
        Listing~\ref{format} shows an example that uses variadic arguments to print formatted output, similar to C's \code{sprintf()} function.
        It is based on a function taken from the PHP Zend Engine.
        As arguments, the function expects a format string \code{fmt}, the variadic arguments \code{ap}, and a buffer \code{xbuf} to which the formatted output should be written.
        To use the function, a C programmer has to invoke a macro to set up and tear down the variadic arguments (respectively \code{va\_start()} and \code{va\_end()}).
        Using the \code{va\_arg()} macro, \code{xbuf\_format\_converter()} can then directly access the variadic arguments.
        The example shows how a string can be accessed (format specifier \code{"\%s"}) that is then inserted into the buffer \code{xbuf}.

        The function uses the format string to determine how many variadic arguments should be accessed.
        For example, for a format string \code{"\%s \%s"} the function attempts to access two variadic arguments that are assumed to have a string type.
        Accessing a variadic argument via \code{va\_arg()} usually manipulates a pointer to the stack and pops the number of bytes that correspond to the specified data type (\code{char *} in our example).
        In a so-called \emph{format string attack}, in which the function reads or writes beyond the stack due to nonexistent arguments, an attacker can exploit the inability of the function to verify the number and the types of the variadic arguments passed~\cite{formatguard,formattype}.

        In CVE-2015-8617, this function was the sink of a vulnerability that existed in PHP-7.0.0. 
        The \code{zend\_throw\_error()} function called \code{xbuf\_format\_converter()} with a \code{message} string that was under user control.
        Consequently, an attacker could use format specifiers without matching arguments to read from and write to memory, and thus execute arbitrary code.
        As another example, in CVE-2016-4448 a vulnerability in libxml2 existed because format specifiers from untrusted input were not escaped.

\begin{lstlisting}[language=C, float=tp,label=format,caption={Example usage of variadic functions, taken from the PHP Zend Engine}]
static void xbuf_format_converter(void *xbuf, const char *fmt, va_list ap) {
  char *s = NULL;
  size_t s_len;
  while (*fmt) {
    if (*fmt != '%') {
      INS_CHAR(xbuf, *fmt);
    } else {
      fmt++;
      switch (*fmt) {
        // ...
        case 's':
        s = va_arg(ap, char *);
        s_len = strlen(s);
        break;
        // ...
      }
      INS_STRING(xbuf, s, s_len);
    }
  }
}
\end{lstlisting}

    }
    \item[Lack of type safety.]{Due to the lack of type safety, a programmer cannot verify whether an object referenced by a pointer corresponds to its expected type~\cite{libcrunch}.
        Listing~\ref{apply} demonstrates this for function pointers.
        The \code{apply()} function expects a function pointer that accepts and returns an \code{int}.
        It uses the function to transform all elements of an array.
        However, its callee might pass a function that returns a double; a call on it would result in undefined behavior.
        Such ``type confusion'' cannot be avoided when calling a function pointer, since objects have no types attached that could be used for validation.
\begin{lstlisting}[language=C, float=tp,label=apply,caption=Example of type confusion]
int apply(int* arr, size_t n, int f(int arg1)) {
  if (f == NULL) return -1;
  for (size_t i = 0; i < n; i++)
    arr[i] = f(arr[i]);
  return 0;
}

double square(int a) { return a * a; }

apply(arr, 5, square);
\end{lstlisting}
    }
    \item[Unterminated strings.]{
        Unterminated strings are a problem, since the string functions of \code{libc} (and sometimes also application code) rely on strings ending with a \code{`\textbackslash0'} (null terminator) character.
        However, C standard library functions that operate on strings lack a common convention on whether to add a null terminator~\cite{strlcpy}.
        Additionally, it is not possible to verify whether a string is properly terminated without potentially causing buffer overreads.
        Listing~\ref{unterminated-string} shows an example of an unterminated string vulnerability.
        The \code{read} function reads a file's contents into a string \code{inputbuf}.
        After the call, \code{inputbuf} is unterminated if the file was unterminated or if \code{MAXLEN} was exceeded.
        This is likely to cause an out-of-bounds write in \code{strcpy()}, since it copies characters to \code{buf} until a null terminator occurs.
        Recent similar real-world vulnerabilities include CVE-2016-7449, where \code{strlcpy()} was used to copy untrusted (potentially unterminated) input in GraphicsMagick.
        Further examples are CVE-2016-5093 and CVE-2016-0055, where strings were not properly terminated in the PHP Zend Engine as well as in Internet Explorer and Microsoft Office Excel~\cite{fortinet}.
    \begin{lstlisting}[language=C, float=tp,label=unterminated-string,caption=Example fragment that may produce and copy an unterminated string]
read(cfgfile, inputbuf, MAXLEN);
char buf[MAXLEN];
strcpy(buf, inputbuf);
puts(buf);
    \end{lstlisting}
    }
    \item[Unsafe functions.]{
        Some functions in common libraries such as \code{libc} have been designed such that they ``can never be guaranteed to work safely''~\cite{cwe242, libsafe}.
        The most prominent example is the \code{gets()} function, which reads user input from \code{stdin} into a buffer passed as an argument.
        Since \code{gets()} lacks a parameter for the size of the supplied buffer, it cannot perform any bounds checking and overflows the buffer if the user input is too large.
        Although C11 replaced \code{gets()} with the more robust \code{gets\_s()} function, legacy code might still require the unsafe \code{gets()} function.
        In general, functions that lack size arguments -- which prevents safe access to arrays -- cannot be made safe without breaking source and binary compatibility.
    }
\end{description}

\section{Introspection Functions}
\label{sec:api}
To enable C programmers to validate arguments and global data, we devised introspection functions to query properties of C objects and the current function (see Appendix~\ref{appendix}).
These functions allow programmers only to inspect objects and not to manipulate them; therefore, the presented functions are not a full reflection interface.

We designed these functions specifically to provide users with the ability to prevent buffer overflow, use-after-free, and other common errors specific to C.
Through introspection, programmers can validate certain properties (memory location, bounds, and types) before performing an operation on an object.
Additionally, introspection allows the number of variadic arguments passed to be queried and their types to be validated.

We built introspection based on several \emph{introspection primitives}.
These primitives are a minimal set of C functions that require run-time support.
We also designed \emph{introspection composites}, which are implemented as normal C functions and are based on the introspection primitives or on other composites.
The introspection functions that we expose to the programmer contain both selected primitives and composites.
We hereafter denote internal functions that are private to an implementation with an underscore prefix.

\subsection{Object Bounds}
\label{sec:size}

Most importantly, we provide functions that enable the programmer to perform bounds checks before accessing an object.
Simply providing a function that returns the size of an object is insufficient, since a pointer can point to the middle of an object.
Instead, we require the runtime to provide two functions to return the space (in bytes) to the left and to the right of a pointer target: \code{\_size\_left()} and \code{\_size\_right()}.
Their result is only defined for \emph{legal} pointers, which we define as pointers that point to valid objects (not \code{INVALID}, see Section~\ref{sec:memory}).

\begin{figure}[tb]
    \centering
    \includegraphics[width=\textwidth*2/5]{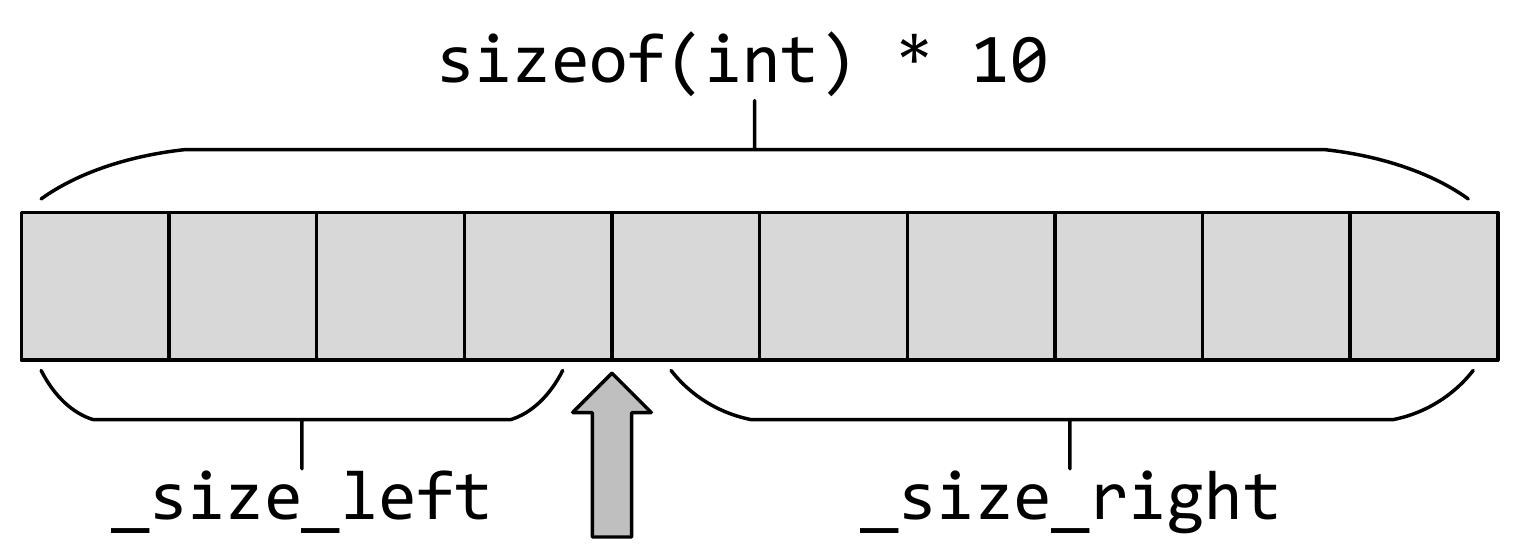}
    \caption{Memory Layout of the Example in Listing~\ref{usage-size-api}}
    \label{usage-size-api-memory-layout}
\end{figure}

Listing~\ref{usage-size-api} illustrates the function return values when passing a pointer to the middle of an integer array to these functions.
For the pointer to the fourth element of the ten-element integer array, \code{\_size\_left()} returns \code{sizeof(int) * 4}, and \code{\_size\_right()} returns \code{sizeof(int) * 6}.
Figure~\ref{usage-size-api-memory-layout} shows the corresponding memory layout.
On an architecture where an \code{int} is four bytes in size the functions return \code{16} and \code{24}, respectively.

\begin{lstlisting}[language=C, float=tp,label=usage-size-api,caption=Example of how to query the space to the left and to the right of a pointee]
int *arr = malloc(sizeof(int) * 10);
int *ptr = &(arr[4]);
printf("%ld\n", size_left(ptr)); // prints 16
printf("%ld\n", size_right(ptr)); // prints 24
\end{lstlisting}

We do not expose these two functions to the programmer, but base the composite functions \code{size\_left()} and \code{size\_right()} on them, which return \code{-1} if the passed argument is not a legal pointer or out of bounds.
Listing~\ref{size-right} shows the implementation of \code{size\_left()}.
Using \code{location()}, the function first checks that the pointer is legal (see Section~\ref{sec:memory}).
It then checks that the spaces to the left and to the right of the pointer are not negative, that is, the pointer is in bounds.
If both checks are passed, the function returns the space to the left of the pointer using \code{\_size\_left()}; otherwise, it returns \code{-1}.
\begin{lstlisting}[language=C, float=tp,label=size-right,caption={Implementation of \code{size\_left()} using the functions \code{location()}, \code{\_size\_left()}, and \code{\_size\_right()}}]
long size_left(const void *ptr) {
  if (location(ptr) == INVALID) return -1;
  bool inBounds = _size_right(ptr) >= 0 && _size_left(ptr) >= 0;
  if (!inBounds) return -1;
  return _size_left(ptr);
}
\end{lstlisting}

Listing~\ref{reduce-function-bounds-check} shows how using \code{size\_right()} improves \code{read\_number()}'s robustness (see Listing~\ref{out-of-bounds}):
If \code{arr} is a valid pointer but points to memory that cannot hold \code{length char}s, we can prevent the out-of-bounds access by aborting the program.
Note that the check also detects lurking bugs, since it aborts even if fewer than \code{length} characters are read.
If \code{arr} is not a valid pointer, the return value of \code{size\_right()} is \code{-1}.

\begin{lstlisting}[language=C, float=tp,label=reduce-function-bounds-check,caption=By using the \code{size\_right()} function we can avoid out-of-bounds accesses in \code{read\_number()}]
void read_number(char* arr, size_t length) {
  int i = 0;
  if (length == 0) return;
  if (size_right(arr) < length) abort();
  // ...
}
\end{lstlisting}

\subsection{Memory Location}
\label{sec:memory}
Querying the memory location of an object (e.g., stack, heap, global data) allows a programmer to obtain information about the lifetime of an object.
For example, it enables programmers to prevent use-after-free errors by detecting whether an object has already been freed.
Another use case is validating that no stack memory is returned by a function.
A programmer can also check whether a location refers to dynamically allocated memory to ensure that \code{free()} can be safely called on it.
For this purpose, we provide the function \code{location()}, which determines where an object lies in memory.

The function returns one of the following enum constants:
\begin{itemize}[leftmargin=10pt]
    \item{\code{INVALID} locations denote \code{NULL} pointers or deallocated memory (freed heap memory or dead stack variables). Programs must not access such objects.}
    \item{\code{AUTOMATIC} locations denote non-static stack allocations. Functions must not return allocated stack variables that were declared in their scope, since they become \code{INVALID} when the function returns. Further, stack variables must not be freed.}
    \item{\code{DYNAMIC} locations denote dynamically allocated heap memory created by \code{malloc()}, \code{realloc()}, or \code{calloc()}. Only memory allocated by these functions can be freed.}
    \item{\code{STATIC} locations denote statically allocated memory such as global variables, string constants, and static local variables. Static compilers usually place such memory in the text or data section of an executable. Programs must not free statically allocated memory.}
\end{itemize}
Listing~\ref{location-example} shows how differently allocated memory relates to the enum constants used by \code{location()}.

\begin{lstlisting}[language=C, float=tp,label=location-example, caption=Example of how the \code{location()} enum constants relate to objects in a program]
int a;                      // location(&a) returns STATIC for global objects
void func() {
  static int b;             // location(&b) returns STATIC for static local objects
  int c;                    // location(&c) returns AUTOMATIC for stack objects
  int* d = malloc(sizeof(int) * 10);
                            // location(&d) returns DYNAMIC for heap objects
  free(d);                  // location(&d) returns INVALID for freed objects
}
\end{lstlisting}

\begin{lstlisting}[language=C, float=tp,label=freeable,caption=By using \code{location()} and \code{\_size\_left()} we can check whether an object can be freed]
bool freeable(const void *ptr) {
  return location(ptr) == DYNAMIC && _size_left(ptr) == 0;
}
\end{lstlisting}

We provide the function \code{freeable()}, which is based on \code{location()}, to conveniently check whether an allocation can be freed.
As Listing~\ref{freeable} demonstrates, a freeable object's location must be \code{DYNAMIC}, and its pointer must point to the beginning of an object.
Listing~\ref{double-free-improved} shows how we can use the \code{freeable()} function to improve the robustness of the code fragment shown in Listing~\ref{use-after-free}.
It ensures that freeing the pointee is valid, and thus prevents invalid free errors, such as double freeing of memory.
Nonetheless, the \code{logError()} function may receive a dangling pointer as an argument.
To resolve this, we can check in \code{logError()} whether the pointer is valid (see Listing~\ref{use-after-free-improved}).

\begin{lstlisting}[language=C, float=tp,label=double-free-improved,caption=By using the \code{freeable()} function we can avoid double-free errors]
char* ptr = (char*) malloc(SIZE * sizeof(char));
if (err) {
  abrt = 1;
  if (freeable(ptr)) free(ptr);
}
// ...
if (abrt) {
  logError("operation aborted", ptr);
  if (freeable(ptr)) free(ptr);
}
\end{lstlisting}

\begin{lstlisting}[language=C, float=tp,label=use-after-free-improved,caption=By using the \code{location()} function we can avoid use-after-free errors]
void logError(const char* message, void* ptr) {
  if (location(ptr) == INVALID)
    log("dangling pointer passed to logError!");
  else
    logf("error while processing %p", ptr);
}
\end{lstlisting}

Note that some libraries, such as OpenSSL, use custom allocators to manage their memory.
Custom allocators are outside the scope of this paper, but could be supported by providing source-code annotations for allocation and free functions; this information could then be used by the runtime to track the memory.
The annotations for the allocation functions would need to specify how to compute the size of the allocated object, and the location of the allocated memory.
Additionally, it might be desirable to add further enum constants, for example, for shared, file-backed, or protected memory.
We omitted additional constants for simplicity.

\subsection{Type}
\label{sec:type}
We provide a function that allows the programmer to validate whether an object is \emph{compatible with} (can be treated as being of) a certain type.
Such a function enables programmers to check whether a function pointer actually points to a function object (and not to a \code{long}, for example) and whether it has the expected function signature.
As another example, programmers can use the function as an alternative to \code{size\_right()} and \code{size\_left()} to verify that a pointer of a certain type can be dereferenced.

C has only a weak notion of types, which makes it difficult to design expressive type introspection functions.
For example, it is ambiguous whether a pointer of type \code{int*} that points to the middle of an integer array should be considered as a pointer to an integer or as a pointer to an integer array.
Another example is heap memory, which lacks a dynamic type; although programmers usually coerce them to the desired type, objects of different types can be stored.
Even worse, when writing to memory, objects can be partially overwritten; for instance, half of a function pointer can be overwritten with an integer value, which makes it difficult to decide whether the pointer is still a valid function pointer.

Instead of assuming that a memory region has a specific type, we designed a function that allows the programmer to check whether the memory region is compatible with a certain type (similar to~\cite{libcrunch}).
The \code{try\_cast()} function expects a pointer to an object as the first argument and tries to cast it to the \code{Type} specified by the second argument.
If the runtime determines that the cast is possible, it returns the passed pointer, and \code{NULL} otherwise.
The cast is only possible if the object can be read, written to, or called as the specified type.

The \code{Type} object is a recursive struct which makes it possible to describe nested types (known as type expressions~\cite{compilers}).
For example, a function pointer with an \code{int} parameter and \code{double} as the return type can be represented by a tree of three \code{Type} structs.
The root struct specifies a \code{function} type and references a struct with an \code{int} type as the argument type as well as a struct with a \code{double} type as the return type.
Since manually constructing \code{Type} structs is tedious, we specified the \emph{optional} operator \code{type()}.
As an argument, it requires the expression \emph{example value}, whose declared type is returned as a \code{Type} run-time data structure.
Since the declared type is a compile-time property, we want to resolve the \code{type()} operator during compile time; consequently, the programmer cannot take \code{type()}'s address and call it indirectly.
The operator is similar to the GNU C extension \code{typeof}, which yields a type that can be used directly in variable declarations or casts.

Listing~\ref{try-cast-reduce} shows how the type introspection functions make the function \code{apply()} (see Listing~\ref{apply}) more robust:
\code{apply()} uses \code{try\_cast()} to check whether the runtime can treat its first argument as the specified function pointer.
Its second argument is the \code{Type} object that the \code{type} operator constructs from the declared function pointer type.
The \code{try\_cast()} function returns the first argument if it is compatible with the specified function pointer type; otherwise, it returns \code{NULL}.
In addition to preventing the calling of invalid function pointers, \code{apply()} prevents out-of-bounds accesses by validating the array size.

\begin{lstlisting}[language=C, float=tp,label=try-cast-reduce,caption=By using \code{try\_cast()} we can ensure that we can perform an indirect call on the function pointer in \code{apply()}]
int apply(int* arr, size_t n, int f(int arg1)) {
  if (size_right(arr) < sizeof(int) * n || try_cast(&f, type(f)) == NULL)
    return -1;
  for (size_t i = 0; i < n; i++)
    arr[i] = f(arr[i]);
  return 0;
}
\end{lstlisting}

The \code{try\_cast()} function is similar to C++'s \code{dynamic\_cast()}.
However, we want to point out that C++'s \code{dynamic\_cast()} works only for class checks (which are well-defined), while our approach works for all C objects.
We believe that the exact semantics of \code{try\_cast()} should be implementation-defined, since run-time information could differ between implementations.
For example, depending on the runtime's knowledge of data execution prevention, it might either allow or reject the cast of a non-executable \code{char} array filled with machine instructions to a function pointer.
Further, different use cases exist, and a security-focused runtime might have more sources of run-time information and be more restrictive than a performance-focused runtime.
For example, a traditional runtime would (for compatibility) allow dereferencing a hand-crafted pointer as long as it corresponds to the address of an object, while a security-focused runtime could disallow it.
Thus, depending on the underlying runtime, compiler, and ABI, \code{try\_cast()} can return different results.

\subsection{Variadic Arguments}

Our introspection interface provides macros to query the number of variadic arguments and enables programmers to access them in a type-safe way.
They are implemented as macros and not as functions, since they need to access the current function's variadic arguments.
The introspection macros make using variadic functions more robust and are, for example, effective in preventing format string attacks~\cite{formatguard}.

Querying the number of variadic arguments can be achieved by calling \code{count\_varargs()}.
The standard \code{va\_arg()} macro reads values from the stack while assuming that they correspond to the user-specified type.
As a robust alternative, introspection composites can use \code{\_get\_vararg()} to access the passed variadic arguments directly by an argument index.
To access the variadic arguments in a type-safe way, we introduced the \code{get\_vararg()} macro, which is exposed to the programmer and expects a type that it uses to call \code{try\_cast()}.
Listing~\ref{avg-improved} shows an example of a function that computes the average of \code{int} arguments.
It uses \code{count\_varargs()} to verify the number of variadic arguments and ensures that the \code{i\textsuperscript{th}} argument is in fact an \code{int} by calling \code{get\_vararg()} with \code{type(\&sum)}.
If an unexpected number of parameters or an object with an unexpected type is passed, the function returns \code{0}.

\begin{lstlisting}[language=C, float=tp,label=avg-improved,caption=By using \code{count\_varargs()} and \code{get\_varargs()} we can use variadics in a robust way]
double avg(int count, ...) {
  if (count == 0 || count != count_varargs())
    return 0;
  int sum = 0;
  for (int i = 0; i < count; i++) {
    int *arg = get_vararg(i, type(&sum));
    if (arg == NULL) return 0;
    else             sum += *arg;
  }
  return (double) sum / count;
}
\end{lstlisting}

For backwards compatibility, we used the introspection intrinsics to make the standard vararg macros (\code{va\_start()}, \code{va\_arg()}, and \code{va\_end()}) more robust.
Firstly, \code{va\_start()} initializes the retrieval of variadic arguments.
We modified it such that it allocates a struct (using the \code{alloca()} stack allocation function) and populates it using \code{\_get\_vararg()} and \code{count\_varargs()}.
The struct comprises the number of variadic arguments, an array of addresses to the variadic arguments, and a counter to index them.
Secondly, \code{va\_arg()} retrieves the next variadic argument.
We modified it such that it checks that the counter does not exceed the number of arguments, increments the counter, indexes the array, and casts the variadic argument to the specified type using \code{try\_cast()}.
If the cast succeeds, the argument is returned; otherwise a call to \code{abort()} exits the program.
Finally, \code{va\_end()} performs a cleanup of the data initialized by \code{va\_start()}.
We modified it such that it resets the variadic arguments counter.

Using the enhanced vararg macros improves the robustness of the \sloppy{\code{xbuf\_format\_converter()}} function (see Listing~\ref{format}), since the number of format specifiers must match the number of arguments, thus making it impossible to exploit the function through format string attacks.
Note that the modified standard macros abort when they process invalid types or an invalid number of arguments, whereas the intrinsic functions allow programmers to react to invalid arguments in other ways.

\section{Implementation}
\label{sec:implementation}
We implemented the introspection primitives in Safe Sulong~\cite{ecoopds}, which is an execution system and bug-finding tool for low-level languages such as C.
At its core is an interpreter written in Java that runs on top of the JVM.
Although this setup is not typical for running C, it is a good experimentation platform because the JVM (and thus also Safe Sulong) already maintains all the run-time metadata that we want to expose.
If exposing introspection primitives turns out to be useful for Safe Sulong, similar mechanisms could also be implemented for other runtimes (e.g., those of static compilation approaches).
Unlike its counterpart Native Sulong~\cite{native-sulong}, Safe Sulong uses Java objects to represent C objects.
By relying on Java's bounds and type checks, Safe Sulong efficiently and reliably detects out-of-bounds accesses, use-after-free, and invalid free.
When detecting such an invalid action, it aborts execution of the program.
Section~\ref{sec:overview} gives an overview of the system, and Section~\ref{sec:concreteimpl} describes how we implemented the introspection primitives.

\subsection{System Overview}
\label{sec:overview}

Figure~\ref{architecture} shows the architecture of Safe Sulong, which comprises the following components:

\begin{figure}[tb]
    \centering
    \includegraphics[width=\textwidth*3/8]{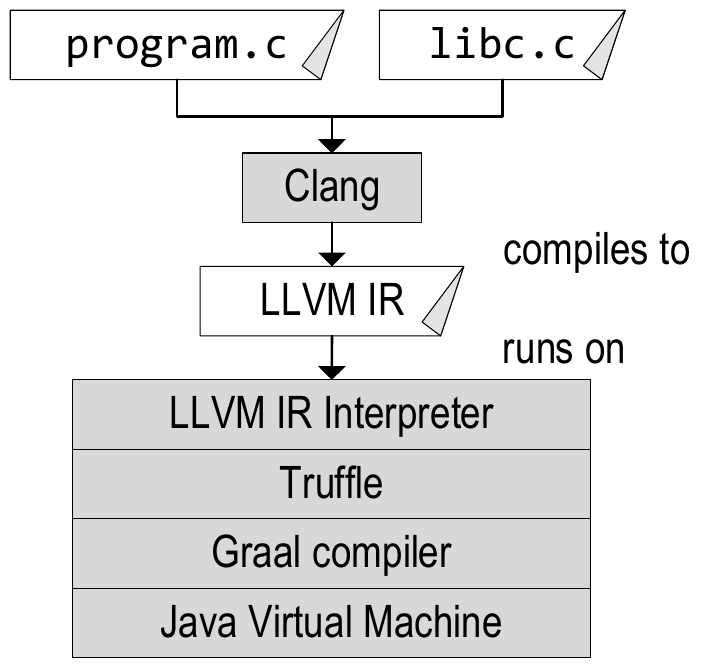}
    \caption{Overview of Safe Sulong}
    \label{architecture}
\end{figure}

\begin{description}[leftmargin=0pt]
    \item[Clang.]{Safe Sulong executes LLVM Intermediate Representation (IR), which represents C functions in a simpler, but lower-level format.
        LLVM is a flexible compilation infrastructure~\cite{llvm}, and we use LLVM's front end Clang to compile the source code (libraries and the user application) to the IR.
    }
    \item[LLVM IR.]{
        LLVM IR retains all C characteristics that are important for the content of this paper.
        It can, for instance, contain external function definitions and function calls.
        By executing LLVM IR, Safe Sulong can execute all languages that can be compiled to this IR, including C++ and Fortran.
        Using binary translators that convert binary code to LLVM IR even allows programs to be executed without access to their source code.
        For example, MC-Semantics~\cite{mcsema} and QEMU~\cite{qemu} support x86, and LLBT~\cite{llbt} supports the translation of ARM code.
        Binary libraries that are converted to LLVM IR can then profit from the enhanced libraries that Safe Sulong can execute, such as our enhanced \code{libc}.
    }
    \item[Truffle.]{The LLVM IR interpreter is based on Truffle~\cite{truffle}.
    Truffle is a language implementation framework written in Java.
    To implement a language, a programmer writes an Abstract Syntax Tree (AST) interpreter in which each operation is implemented as an executable node.
    Nodes can have children that parent nodes can execute to compute their results.
    }
    \item[Graal.]{Truffle uses Graal~\cite{rulethemall}, a dynamic compiler, to compile frequently executed Truffle ASTs to machine code.
        Graal applies aggressive optimistic optimizations based on assumptions that are later checked in the machine code.
        If an assumption no longer holds, the compiled code \emph{deoptimizes}~\cite{deoptimization}, that is, control is transferred back to the interpreter and the machine code of the AST is discarded.
    }
    \item[LLVM IR Interpreter.]{The LLVM IR interpreter forms the core of Safe Sulong; it executes both the user application and the enhanced \code{libc}.
        First, a front end parses the LLVM IR and constructs a Truffle AST for each LLVM IR function.
        Then, the interpreter starts executing the \code{main} function AST, which can invoke other ASTs.
        During execution, Graal compiles frequently executed functions to machine code.
    }
    \item[JVM.]{The system can run efficiently on any JVM that implements the Java-based JVM compiler interface (JVMCI~\cite{jvmci}).
        JVMCI supports Graal and other compilers written in Java.
    }
\end{description}
\begin{figure}
    \centering
    \includegraphics[width=\textwidth*3/5]{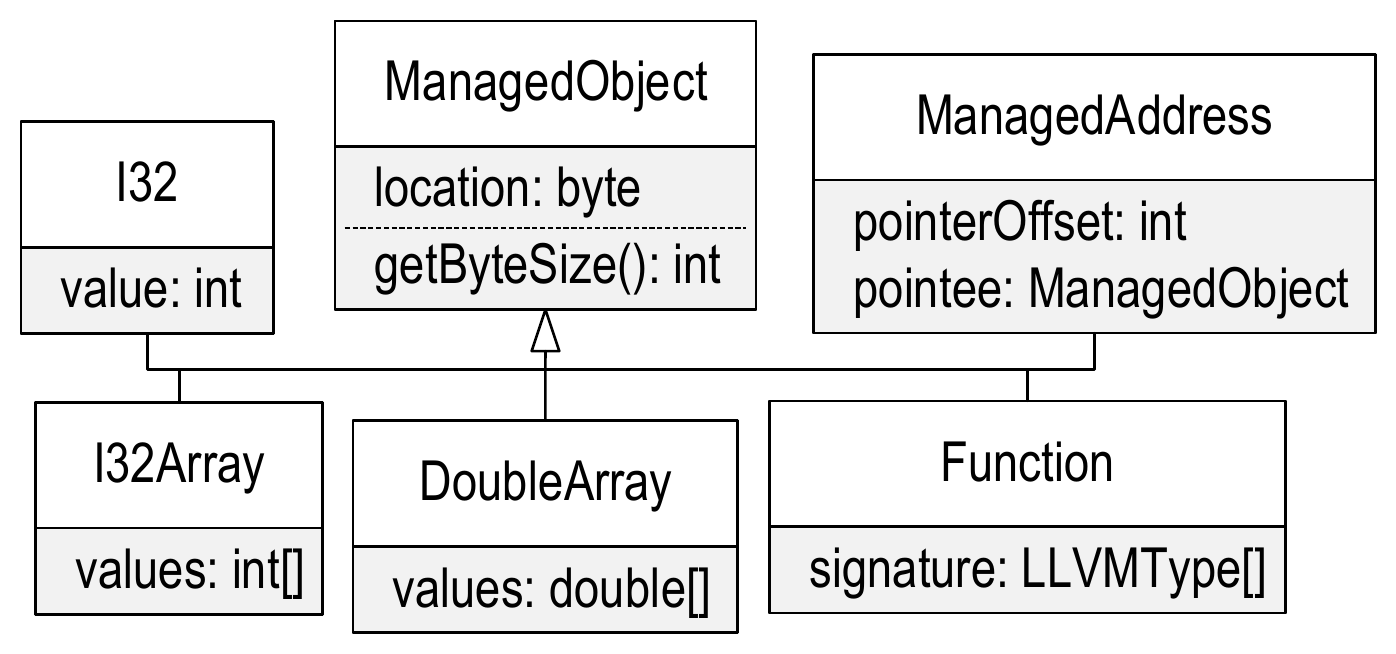}
    \caption{Diagram of the \code{ManagedObject} Hierarchy}
    \label{classes}
\end{figure}

\subsection{Introspection Primitives and Other Functions}
While the majority of Safe Sulong's \code{libc} is implemented in C, the introspection primitives (and a core API, similar to system calls) are implemented directly in Java.
Both are ultimately represented using executable ASTs, which are stored in a symbol table created prior to program execution.
For functions contained in the LLVM IR file, the parser constructs the AST nodes from the instructions denoted in the LLVM IR function.
For introspection primitives, we implemented special nodes that have no equivalent bitcode instruction (see Section~\ref{sec:concreteimpl}).
During execution, Safe Sulong looks up the AST in the symbol table using the function name.
From the runtime's perspective, the implementation of that function is transparent.

\subsection{Objects and Introspection}
\label{sec:concreteimpl}

The LLVM IR interpreter uses Java objects instead of native memory to represent LLVM IR objects (and thus C objects).
Figure~\ref{classes} illustrates its type hierarchy.
Every LLVM IR object is a \code{ManagedObject} which has subclasses for the different types.
For example, an \code{int} is represented by an \code{I32} object, which stores the \code{int}'s value in the \code{value} field.
Similarly, there are subclasses for arrays, functions, pointers, structs, and other types.
Note that we have previously described a similar object hierarchy for the implementation of a \emph{Lenient C} dialect and how certain corner cases are supported (e.g., deriving pointers from integers)~\cite{lenientc}.
In the introspection implementation, we needed to expose properties of these Java objects to the programmer:

\begin{description}[leftmargin=0pt]
    \item[Bounds.]{
        The \code{ManagedObject} class provides the method \code{getByteSize()}, which returns the size of an object.
        Safe Sulong represents pointers as objects of a \code{ManagedAddress} class that holds a reference to the pointee and a pointer offset that is updated through pointer arithmetics (\code{pointee} and \code{pointerOffset}).
        For example, for the pointer to the \code{4\textsuperscript{th}} element of an integer array in Listing~\ref{usage-size-api}, the \code{pointerOffset} is 16, and \code{pointee} references an \code{I32Array} that holds a Java \code{int} array (see Figure~\ref{class-size-example}).
        If a program were to dereference the pointer, the interpreter would compute \code{pointerOffset / sizeof(int)} to index the array.
        We implemented the \code{size\_right()} function by \code{ptr.pointee.getByteSize() - ptr.pointerOffset}.
    }
    \item[Memory location.]{
        Although \code{ManagedObject}s live on the Java heap, the \code{location()} function needs to return their \emph{logical} memory location.
        This location is stored in a field of the \code{ManagedObject} class.
        Depending on whether an object is allocated through \code{malloc()}, as a global variable, as a static local variable, or as a constant, we assign a different flag to its \code{location} field; calls to \code{free()} and deallocation of automatic variables assign \code{INVALID}.
        For instance, for an integer array that lives on the stack, the interpreter allocates an \code{I32Array} and assigns \code{AUTOMATIC} to its \code{location}.
        After leaving the function scope, its \code{location} is updated to \code{INVALID}.
        When the \code{location()} function is called with a pointer to the integer array, it returns the \code{location} field's value.
    }
    \item[Type.]{
        For implementing the \code{try\_cast()} function, we check if the type of the passed object (given by its Java class) is compatible with the type specified by the \code{Type} struct.
        For example, to check whether we can call a pointer as a function with a certain signature, we first compare the passed pointer with a \code{Type} that describes this signature.
        If the pointer references a Safe Sulong object of type \code{Function}, the argument and return types are compared.
        This is possible because \code{Function} objects retain run-time information about their arguments and return types, which can be retrieved via the method \code{getSignature()}.
    }
    \item[Variadic arguments.]{
        In Safe Sulong, a caller explicitly passes its arguments as an object array (i.e., a Java array of known length) to its callee.
        Based on the function signature and the object array, the callee can count the variadic arguments to implement \code{count\_varargs()} and extract them to implement \code{\_get\_vararg()}.
    }
\end{description}

\section{Case Study: Safe Sulong's Standard Library}
\label{sec:casestudy}

We implemented an enhanced \code{libc} for Safe Sulong.
This \code{libc} uses introspection for checks that make it more robust against usage errors and attacks.
For instance, its functions identify invalid parameters that would otherwise cause out-of-bounds accesses or use-after-frees.
In such a case, the functions return special values to indicate that something went wrong, and then set \code{errno} to an error code.
However, for functions for which no special value can be returned (e.g., because the return type is \code{void}), setting \code{errno} would be meaningless, since functions are allowed to change \code{errno} arbitrarily even if no error occurred.
In these cases, the functions still attempt to compute a meaningful result.
Such behavior is compliant with the C standards, since we prevent illegal actions with undefined behavior that could crash the program or corrupt memory.

For applications and libraries that run on Safe Sulong, the distribution format is LLVM IR and not executable code.
Our standard library improvements are binary-compatible at the IR level, which means that users do not have to recompile their applications when using our enhanced \code{libc}.
In addition, this standard library is source-compatible, so a user is not required to change the program when using it.
Below, we give an overview of our enhanced library functions:
\begin{figure}[tb]
    \centering
    \includegraphics[width=\textwidth*3/5]{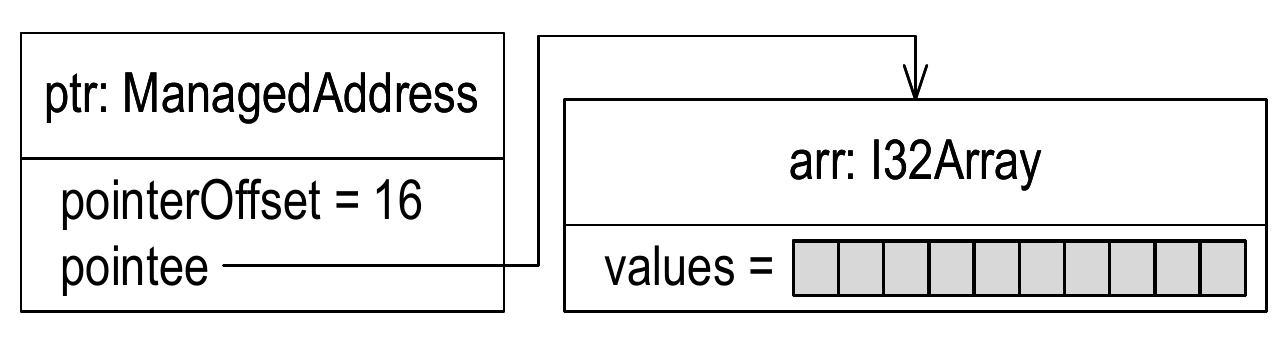}
    \caption{Representation of a pointer to the \code{4\textsuperscript{th}} element of an int array}
    \label{class-size-example}
\end{figure}

\begin{description}[leftmargin=0pt]
    \item[String functions.]{We made all functions that operate on strings (\code{strlen()}, \code{atoi()}, \code{strcmp()}, \code{printf()}, etc.) more robust by computing meaningful results even when a string lacks a null terminator.
        They do not read or write outside the boundaries of unterminated strings, which makes them robust against common string vulnerabilities.
        The functions increase availability of the system, since unterminated strings passed to \code{libc} do not cause crashes.
        Note that when a function outside \code{libc} relies on a terminated string, it will still trigger an out-of-bounds access and cause Safe Sulong to abort execution.
        Thus, increased availability does not harm confidentiality (e.g., by leaking data of other objects) and integrity (e.g., by overwriting other objects).

        For instance, Listing~\ref{saferstrlen} shows how we improved \code{strlen()} by preventing buffer overflows when iterating over a string, and by improving the handling of non-legal pointers (where \code{size\_right()} returns \code{-1}).
        For terminated strings, \code{strlen()} iterates until the first \code{`\textbackslash0'} character to return the length of the string.
        For unterminated strings, the function cannot return \code{-1} to indicate an error, since \code{size\_t} is unsigned, so we also do not set \code{errno}.
        Instead, it iterates until the end of the buffer and returns the size of the string until the end of the buffer.

\begin{lstlisting}[language=C, float=tp,label=saferstrlen,caption={Robust implementation of \code{strlen()} that also works for unterminated strings}]
size_t strlen(const char *str) {
  size_t len = 0;
  while (size_right(str) > 0 && *str != '\0') {
    len++; str++;
  }
  return len;
}
\end{lstlisting}
    
    The enhanced string functions also allow execution of the code fragment in Listing~\ref{unterminated-string}.
    Even though the source string may be unterminated, \code{strcpy()} will not produce an out-of-bounds read, since it stops copying when reaching the end of the source or destination buffer.
    The call to \code{puts()} also works as expected, and prints the unterminated string.
    }
    \item[Functions that free memory.]{We made functions that free memory (\code{realloc()} and \code{free()}) more robust by checking whether their argument can safely be freed using \code{freeable()}.
        In Safe Sulong, \code{malloc()} is written in Java and allocates a Java object.
        By using the introspection functions we were able to conveniently and robustly implement \code{realloc()} in C without having to maintain a list of allocated and freed objects.
    }
    \item[Format string functions.]{
        We made input and output functions that expect format strings more robust.
        Examples are the \code{printf()} functions (\code{printf()}, \code{fprintf()}, \code{sprintf()}, \code{vfprintf()}, \code{vprintf()}, \code{vsnprintf()}, \code{vsprintf()}) and the \code{scanf()} functions (\code{scanf()}, \code{fscanf()}, etc.).
        These functions expect format strings that contain format specifiers, and matching arguments that are used to produce the formatted output.
        Since the functions are variadic, we used \code{count\_varargs()} to add checks that verify that the number of format specifiers is equal to the actual number of arguments.
        Further, the functions use \code{get\_vararg()} to verify the argument types.
        This prevents format-string vulnerabilities and out-of-bounds reads in the format string, as demonstrated in the implementation of \code{strlen()}.
    }
    \item[Higher-order functions.]{We enhanced functions that receive function pointers such as \code{qsort()} and \code{bsearch()}.
        Listing~\ref{saferqsort} shows how \code{qsort()} can use \code{try\_cast()} to verify that \code{f} is a function pointer that is compatible with the specified signature.
        Furthermore, the functions verify that no memory errors, such as buffer overflows, can occur.
\begin{lstlisting}[language=C, float=tp,label=saferqsort,caption={Robust \code{qsort()} implementation that checks whether it can call the supplied function pointer}]
void qsort(void *base, size_t nitems, size_t size, int (*f)(const void *, const void*)) {
  int (*verifiedPointer)(const void *, const void*) = try_cast(&f, type(f));
  if (size_right(base) < nitems * size || verifiedPointer == NULL) errno = EINVAL;
  else {
    // qsort implementation
  }
}
\end{lstlisting}
    }
    \item[gets() and gets\_s().]{
        While C11 replaced the \code{gets()} function with \code{gets\_s()}, Safe Sulong can still provide a robust implementation for \code{gets()} (see Listing~\ref{safegets}).
        Since \code{size\_right()} can determine the size of the buffer to the right of the pointer, we can call it and use the returned size as an argument to the more robust \code{gets\_s()} function.
        If the pointer is not legal, we pass \code{0}, which \code{gets\_s()} handles as an error.
    We also made \code{gets\_s()} more robust against erroneous parameters (see Listing~\ref{safergetss}).
        By using \code{size\_right()} we can validate that the size parameter \code{n} is at least as large as the remaining space right of the pointer.
        The check prevents buffer overflows for \code{gets()} and \code{gets\_s()}, and also passing of dead stack memory or freed heap memory.
\begin{lstlisting}[language=C, float=tp,label=safegets,caption={Robust implementation of \code{gets()} that uses the more robust \code{gets\_s()} in its implementation}]
char *gets(char *str) {
  int size = size_right(str);
  return gets_s(str, size == -1 ? 0 : size);
}
\end{lstlisting}
\begin{lstlisting}[language=C, float=tp,label=safergetss,caption=Robust implementation of \code{gets\_s()} that verifies the passed size argument]
char *gets_s(char *str, rsize_t n) {
  if (size_right(str) < (long) n) {
    errno = EINVAL; return NULL;
  } else {
    // original code
  }
}
\end{lstlisting}
    }
\end{description}

\section{Related Work}

\begin{description}[leftmargin=0pt]
    \item[C Memory safety approaches.]{
        For decades, academia and industry have been coming up with approaches to tackling memory errors in C.
        Thus, there is a vast number of approaches that deal with these issues, both static and run-time approaches, both hardware- and software-based.
        We consider our approach as a run-time approach, since the checks (specified by programmers in their programs) are executed during run time.
        The literature provides a historical overview of memory errors and defense mechanisms~\cite{pastpresentfuture}, an investigation of the weaknesses of current memory defense mechanisms including a general model for memory attacks~\cite{sok}, and a survey of vulnerabilities and run-time countermeasures~\cite{runtime}.
        Using introspection to prevent memory errors is a novel approach that is complementary to existing approaches because the programmer can check for and prevent an invalid action; if the check is omitted and an invalid access occurs, an existing memory safety solution could still prevent the access.
    }
    \item[Run-time types for C.]{
        \code{libcrunch}~\cite{libcrunch} is a system that detects type-cast errors at run time.
        It is based on \code{liballocs}~\cite{liballocs}, a run-time system that augments Unix processes with allocation-based types.
        \code{libcrunch} provides an \code{\_\_is\_a()} introspection function that exposes the type of an object.
        It uses this function to validate type casts and issues a warning on unsound casts.
        In contrast to our approach, \code{libcrunch} checks for invalid casts automatically, so the \code{\_\_is\_a()} function is not exposed to the programmer, nor are there other introspection functions.
        However, we believe that the system could be extended to provide additional run-time information that could be used to implement the introspection primitives.
        Typical overheads of collecting and using the type information are between 5-35\%, which demonstrates that introspection functions are feasible in static compilation approaches.
    }
    \item[Failure-oblivious computing.]{
        Failure-oblivious computing~\cite{failureobliviouscomputing} is a technique that enables servers to continue their normal execution path in the presence of memory errors.
        Instead of aborting the program, invalid writes are discarded, and for invalid reads values are manufactured.
        Note that this approach is automatic, since the compiler inserts checks and continuation code where memory errors can occur.
        Failure-oblivious computing would, for example, work well for \code{strlen} by manufacturing the value zero when the \code{NULL} terminator is missing and the read runs over the buffer end.
        However, returning zero for out-of-bounds accesses does not work in general; for example, when the loop's exit condition checks if the array element is \code{-1}, failure-oblivious computing approaches could run into an endless loop.
        In contrast, using our introspection technique, programmers can take into account the semantics of a function to prevent such situations.
        Additionally, introspection can also be used for bug-finding (not only to increase availability), for example, by checking if the actual buffer length corresponds to the expected buffer length in functions like \code{gets\_s}.
    }
    \item[Static vulnerability scanners.]{
        Static vulnerability scanners identify calls to unsafe functions such as \code{gets()} depending on a policy specified in a vulnerability database~\cite{its4}.
        Such approaches must decide conservatively whether a call is allowed, unlike our approach, which validates parameters at run-time through introspection.
        Nowadays, most compilers issue a warning when they identify a call to an unsafe function such as \code{gets()}, but not necessarily for other, slightly safer functions, such as \code{strcpy()}.
    }
    \item[Fault injection to increase library robustness.]{
        Fault injection approaches generate a series of test cases that exercise library functions in an attempt to trigger a crash in them.
        HEALERS~\cite{increase-library-robustness,generatorarchitecture} is an approach that, after identifying a non-robust function, automatically generates a wrapper that sits between the application and its shared libraries to handle or prevent illegal parameters.
        To check the bounds of heap objects passed to the functions, the approach instruments \code{malloc()} and stores bounds information.
        In contrast to our solution, the approaches above support pre-compiled libraries.
        However, they can generate wrapper checks only where run-time information is explicitly available in the program.
        Additionally, they prevent the programmer from specifying the action in case of an error, and always set \emph{errno} and return an error code.
    }
    \item[Detecting API misusages.]{
        APISan~\cite{apisan} is a tool for finding API usage errors, such as cryptographic protocol API misues, but also integer overflows, \code{NULL} dereferences, memory leaks, incorrect return values, format string vulnerabilities, and wrong arguments.
        It is based on the idea that the dominant usage pattern of an API across several projects indicates its correct use.
        APISan is implemented by gathering execution traces using symbolic execution, from which it infers correct API usages; deviating patterns are potential API misuses.
        While this approach aims to identify incorrect use of libraries, our approach aims to make library functions more robust.
    }
    \item[Replacing (parts of) libc.]{
        SFIO~\cite{sfio} is a \code{libc} replacement and addresses several of its problems.
        It mainly improved completeness and efficiency, but it also introduced safer routines for functions that operate on format strings.
        Additionally, the SFIO standard library functions are more consistent in their arguments and argument order, and thus less error-prone than some of the \code{libc} functions.
        In \cite{strlcpy}, the less error-prone \code{strlcpy()} and \code{strlcat()} functions were presented as replacements for the \code{strcpy()} and \code{strncat()} functions.
        Unlike our improved C standard library, these approaches lack source compatibility.
    }

    \item[Safer implementation of library functions.]{
        To prevent format string vulnerabilities in the \code{printf} family of functions, FormatGuard~\cite{formatguard} uses the preprocessor to count the arguments to variadic functions during compile time and checks that the number complies with the actual number at run time.
        FormatGuard replaces the \code{printf} functions in the C standard library with more secure versions while retaining compatibility with most programs.
        From a user perspective, FormatGuard is similar to Safe Sulong's standard library, in that both provide more robust C standard library functions.
        While our approach works only for runtimes that implement the introspection primitives, StackGuard works for arbitrary compilers and runtimes.
        However, our approach can also verify bounds, memory location, and types of objects.
    }
    \item[Restricting buffer overflows in library functions.]{
        Libsafe~\cite{libsafe} replaces calls to unsafe library functions (such as \code{strcpy()} and \code{gets()}) with wrappers that ensure that potential buffer overflows are contained within the current stack frame.
        It can prevent only stack buffer overflows, since it checks that write accesses do not extend beyond the end of the buffer's stack frame.
        In contrast, approaches exist that protect only against heap buffer overflows caused by C standard library functions~\cite{containment-wrappers}.
        By intercepting C standard library calls, the approach keeps track of heap memory allocations and performs bounds checking before calling the C standard library functions that operate on buffers.
        Both approaches work with any existing pre-compiled library, but do not protect against all kinds of buffer overflows.
        With our approach, a programmer can implement checks that prevent both heap and stack overflows, and use the introspection interface to also prevent use-after-free and other errors.
    }
    \item[Reflection for C.]{
        Higher-level languages such as Java or C\# throw exceptions when encountering out-of-bounds accesses and other errors.
        Exception handling is a more expressive approach than explicitly checking for invalid accesses in advance, since it separates the two concerns in the program.
        Some approaches introduced mechanisms to raise and catch exceptions in C~\cite{exceptionalc, exceptionsc}.
        However, these approaches do not describe how invalid memory errors could be caught and exposed to the programmer as an exception.
    }
\end{description}

\section{Discussion}

\begin{description}[leftmargin=0pt]
    \item[Advantages over existing tools.]{
        We assume that introspection is exposed by a runtime that automatically aborts when detecting an error (e.g., an out-of-bounds access).
        In this scenario, using introspection allows programmers to override the default behavior of aborting the program by checking for invalid states and by reacting to them before the failure occurs.
        Even if checks are omitted, the runtime aborts execution in case of an error.
        Additionally, introspection can be used to check for faults that might not result in errors during run time.
        While adding these checks does not come for free (i.e., they require programming effort), we believe that they can be useful at boundaries of shared libraries, and at the boundaries of subcomponents within a project.
    }
    \item[Adoption of introspection.]{
        Two of the C/C++ tenets are that ``you don't pay for what you don't use''~\cite{cpp} and to ``trust the programmer''~\cite{rationalec99}.
        Hence, programmers often eschew checks even if they are possible without introspection functions~\cite{increase-library-robustness}.
        An open question is thus whether C programmers would use introspection if they had access to it.
        We believe that there is a need for the safe execution of legacy C code (at the expense of performance) as an alternative to porting programs to safer languages.

        It has yet to be determined which of the introspection functions are useful in practice (e.g., by conducting a case study on real-world programs).
        We believe that functions such as \code{size\_right()} are easy to understand and use, and could prevent common errors in practice.
        In contrast, grasping the semantics of \code{try\_cast()} is more difficult because C does not have a strong notion of typing, and use cases for it are also rare; consequently, it would probably be used less often.
    }
    \item[Safer languages.]{
        Since using introspection requires changes to the source code, a question is whether a library should not simply be rewritten in some other systems programming language, such as Rust or Go, that approach the performance of C while being safe.
        First, preventing out-of-bounds accesses or use-after-free errors can already be prevented by using special runtimes without rewriting the project in a safer language (e.g., using AddressSanitizer~\cite{asan} or SoftBound+CETS~\cite{softbound,cets}).
        However, our approach goes beyond these guarantees by allowing the programmer to handle errors in customized ways.
        Second, the effort required to rewrite an application would simply be too high for many real-world applications.
        In contrast, incrementally adding checks to an existing code base is less work.
    }
    \item[Legacy code.]{
        Our approach also brings benefits for legacy applications, namely when a commonly used shared library is modified to employ introspection for additional checks:
        For example, there are legacy applications that use the insecure \code{gets()} \code{libc} function.
        Using our approach, a safe implementation of \code{gets()} can be provided if the runtime implements the introspection interface and \code{libc} uses it to query the length of the buffer.
        Thus, availability or security of legacy code can be improved simply by employing a \code{libc} that inserts additional checks enabled by introspection.
        In contrast, when reimplementing \code{libc} in a safer language, the function \code{gets()} cannot be made safe, as a buffer allocated by C code has no bounds information attached to it.
      }
    \item[Static compilation.]{
        Introspection requires information about run-time properties of objects in the program.
        While interpreters and virtual machines often maintain this information, runtimes that execute native programs compiled by static compilers such as Clang or GCC do not.
        We want to point out that debug metadata (obtained by compiling with the \code{-g} flag) cannot provide per-object type information needed for introspection.
        However, it has been shown that per-object information (such as types) can be added add low cost to static compilation approaches~\cite{libcrunch} and hence make implementing the introspection functions in their runtimes feasible.
        As part of future work, we intend to implement introspection primitives using tools based on a static compilation model.
    }
    \item[Partial metadata availability.]{
        While designing the interface, we assumed that a tool that implements introspection maintains all relevant metadata.
        However, some runtimes maintain only a subset of it; for example, bounds checkers track bounds information and can implement only \code{\_size\_left()} and \code{\_size\_right()}.
        Custom memory allocators that track heap allocations can implement only a subset of the function \code{location()}.
        It has yet to be investigated how code can benefit from runtimes that implement only parts of the interface.
        A compile-time approach would involve checking introspection features using preprocessor directives.
        Another approach would involve structuring the checks such that they do not fail when an introspection function returns a default value that indicates that the corresponding feature is unsupported.
    }
    \item[Performance measurement.]{
        The focus of this work was on evaluating the usefulness of exposing introspection functions to library writers.
        We did not invest much time in optimizing the peak performance of our approach in Safe Sulong.
        Thus, we show its performance only on a small set of microbenchmarks for which we used our enhanced \code{libc} (see Appendix~\ref{appendix:performance}).
        As part of future work, we want to extend Safe Sulong's completeness to execute larger benchmarks, such as SPEC INT~\cite{specint}.
    }
\end{description}

\section{Conclusion}

We have presented an introspection interface for C that programmers can use to make libraries more robust.
The introspection functions expose properties of objects (bounds, memory location, and type) as well as properties of variadic functions (number of variadic arguments and their types).
We have described an implementation of the introspection primitives in Safe Sulong, a system that provides memory-safe execution of C code.
However, our approach is not restricted to Safe Sulong; many dynamic bug-finding tools and runtimes exist that could implement (a subset of) the introspection interface.
The approach is complementary to existing memory safety approaches, as programmers can use it to react to and prevent errors in the application logic.
Finally, we have shown how we used the introspection interface to implement an enhanced, source-compatible C standard library.

\acks
We thank the anonymous reviewers for their valuable comments and suggestions to improve the quality of the paper.
We also thank Ingrid Abfalter, whose proofreading and editorial assistance greatly improved the manuscript.
We thank all members of the Virtual Machine Research Group at Oracle Labs and the Institute of System Software at Johannes Kepler University Linz for their support and contributions.

\printbibliography

\appendix

\newpage{}
\section{Preliminary Performance Evaluation}
\label{appendix:performance}
The focus of this work was on evaluating the usefulness of exposing introspection functions to library writers.
We have not yet invested much time in optimizing the peak performance of our approach in Safe Sulong.
To demonstrate that Safe Sulong can run programs in a testing environment, we ran six benchmarks of the Computer Language Benchmark Game~\cite{shootouts} (binarytrees, fannkuchredux, fasta, mandelbrot, nbody, and spectralnorm) and the whetstone benchmark~\cite{whetstone}, once with the enhanced \code{libc} and once without introspection checks.
We determined the average peak performance of 10 runs by measuring the execution time after 50 in-process warm-up iterations.
On these benchmarks, Safe Sulong's peak performance was 2.3$\times$ slower than executables compiled by Clang with all optimizations turned on (\code{-O3} flag).
We were unable to find any observable performance differences between the two \code{libc} versions, which is in part due to some of the introspection checks redundantly duplicating automatic checks performed by the JVM (e.g., bounds checks); such redundant checks can be eliminated by using the Graal compiler (e.g., through conditional elimination~\cite{scala}).
As part of future work, we will evaluate Safe Sulong's performance in combination with the enhanced \code{libc} on larger benchmarks that stress the introspection functionality.

\section{Introspection Functions}
\label{appendix}

Table~\ref{tbl:introspection} shows the functions and macros of the introspection interface.
Internal functions that are private to the implementation are denoted with an underscore prefix.

\begin{table}
\caption{Functions and macros of the introspection interface}
\centering
\begin{tabular}{p{5cm} p{1.8cm} p{7cm}}
\hline
\multicolumn{3}{|c|}{\cellcolor{gray!15}Object bounds functions}\\
\hline
\code{long \_size\_right(void *)} & Primitive internal  & Returns the space in bytes from the pointer target to the end of the pointed object. This function is undefined for illegal pointers.                                  \\
\hline
\code{long \_size\_left(void *)}  & Primitive internal  &    Returns the space in bytes from the pointer target to the beginning of the pointed object. This function is undefined for illegal pointers.                                                                       \\
\hline
\code{long size\_right(void *)}     & Composite          & Returns the remaining space in bytes to the right of the pointer. Returns -1 if the pointer is not legal or out of bounds.                                         \\
\hline
\code{long size\_left(void *)}      & Composite           & Returns the remaining space in bytes to the left of the pointer. Returns -1 if the pointer is not legal or out of bounds.                                       \\
\hline
\multicolumn{3}{|c|}{\cellcolor{gray!15}Memory location functions}\\
\hline
\code{Location location(void *)}     & Primitive           & Returns the kind of the memory location of the referenced object. Returns -1 if the pointer is \code{NULL}.                                                         \\
\hline
\code{bool freeable(void *)}         & Composite           & Returns whether the pointer is freeable (i.e., \code{DYNAMIC} non-null memory; pointer referencing the beginning of an object). \\
\hline
\multicolumn{3}{|c|}{\cellcolor{gray!15}Type functions}\\
\hline
\code{void* try\_cast(void *, struct Type *)} & Primitive & Returns the first argument if the pointer is legal, within bounds, and the referenced object can be treated as of being of the specified type and \code{NULL} otherwise. \\
\hline
\multicolumn{3}{|c|}{\cellcolor{gray!15}Variadic function macros}\\
\hline
\code{int count\_varargs()} & Primitive & Returns the number of variadic arguments that are passed to the currently executing function.\\
\hline
\code{void* \_get\_vararg(int i)} & Primitive internal & Returns the \code{i\textsuperscript{th}} variadic argument (starting from 0) and returns \code{NULL} if \code{i} is greater or equal to \code{count\_varargs()}.\\
\hline
\code{void* get\_vararg(int i, Type* type)} & Composite & Returns the \code{i\textsuperscript{th}} variadic argument (starting from 0) as the specified \code{type}. Returns \code{NULL} if the object cannot be treated as being of the specified type or if \code{i} is greater or equal to \code{count\_varargs()}.\\
\hline
\end{tabular}
\label{tbl:introspection}
\end{table}

\end{document}